# Calculation of Special Spin Behavior of $Dy^{3+}$ in $DyFe_{1-x}Cr_xO_3$ System by Molecular Field Model

Kaiyang Gao[1], Kexuan Zhou[1], Jiyu Shen[1], Zeyi Lu[1], Chenying Gong[1], Zhongjin Wu[1], Ke Shi[1], Jing Guo[1], Zhaoyi Wang[3], Min Liu*[,1,2]

[1]College of Nuclear Science and Technology, University of South China, Hengyang 421200, Hunan, P.R China.
[2]Zhuhai Tsinghua University Research Institute Innovation Center, 101 University Ave, Tangjiawan Zhuhai 519000, P.R China.
[3]College of physics and Electronic Sciences, Hunan Normal University, Changsha 410081, Hunan, P.R China.

Email: liuhart@126.com, xiayfusc@126.com

**Abstract**

In this study, the sol-gel method synthesized the magnetic measurement and analysis of single-phase polycrystalline perovskite $DyFe_{1-x}Cr_xO_3$ (DFCO). The experimental data were fitted and calculated by a four-sublattice molecular field model. Unlike previous studies, we found that in $DyFe_{1-x}Cr_xO_3$, the spin of the A-site rare earth ion $Dy^{3+}$ also changed simultaneously with the spin reorientation of the $Fe^{3+}/Cr^{3+}$ ions. The effective spin is defined as the projection of the A site's total spin on the B site's spin plane, and the curve of temperature changes is obtained after fitting. With this theory, a very accurate thermomagnetic curve is obtained by fitting. This is convincing and, at the same time, provides a reference for the development of spintronic devices in the future.

**Introduction**

Rare earth-based perovskites have always been critical raw materials for information storage devices and read/write heads [1,2] and are also widely studied as magneto-optical effects. [3-6] This is attributed to the rich and complex magnetism of this class of materials; especially after the addition of rare earth ions with a net magnetic moment, the source of their magnetic nature will become very complex. In the past few years, many scholars have done exciting research on it. Wu et al. [7] reported that the twofold spin reorientation behavior of the double rare earth orthoferrite $Dy_{0.5}Pr_{0.5}FeO_3$ originated from the effective anisotropic field generated by the $R^{3+}$ ion spin interaction and could be controlled by an external magnetic field. Park et al. [8] calculated by density functional and dynamical mean field theory that the metal-insulator transition of Rare-earth-element nickelates originates from the site-selective Mott phase. Najee et al. [9] recently demonstrated direct, reversible, and continuous control of the ferromagnetism of rare-earth titanium-based perovskites by uniaxial strain affecting the tilt and rotation of $TiO_6$ octahedra.

$DyFe/CrO_3$ is a typical perovskite whose magnetic properties consist of complex exchange interactions due to the presence of two transition metal ions at the B site: $Dy^{3+}$-$Dy^{3+}$, $Dy^{3+}$-$Fe^{3+}$, $Dy^{3+}$-$Cr^{3+}$, $Fe^{3+}$-$Fe^{3+}$, $Cr^{3+}$-$Cr^{3+}$, $Fe^{3+}$-$Cr^{3+}$. In general, up to the Néel phase transition temperature of $Fe^{3+}$-$Fe^{3+}$/$Cr^{3+}$-$Cr^{3+}$ (approximately 650 K to 750 K for Fe and 120 K to 300 K for Cr) [10.11], the overall magnetism of the system is directly contributed by the G-type antiferromagnetic structure of $Fe^{3+}$/$Cr^{3+}$. At the ultra-low temperature of around 5 K, [12] the magnetic moment of $Dy^{3+}$-$Dy^{3+}$ is fixed to the ordered structure, and a firm antiferromagnetic order appears. Still, above 5 K temperature, it only contributes to a strong paramagnetism.

In most rare earth perovskite, there is a particular spin orientation behavior - spin reorientation. This spintronics phenomenon originates from the strong coupling between the high-spin rare earth ion (4f) and the transition metal ion (3d). In many previous studies, the spin reorientation behavior of B-site ions has been investigated by growing single crystals and attaching external magnetic fields in different easy-axis directions. [13-15] However, very little research has been done on magnetic simulation.

This study uses the four-lattice molecular field theory to perform a magnetic fit to a polycrystalline DFCO system. It is found that not only do the transition metal ions at the B-site undergo a spin orientation transition during thermogenic spin reorientation, but also the A-site rare earth ions undergo a similar continuous spin change simultaneously. Finally, the values of the exchange constants and the effective spins, closely related to the macroscopic magnetism, are obtained. This approach has been studied with excellent results in other partial rare earth perovskite systems. [16-19]

**Experimental**

Sample Preparation

$DyFe_{1-x}Cr_xO_3$ (x=0.1 and 0.9) was prepared by a simple sol–gel combustion method. The required precursors are all from McLean, namely Dysprosium nitrate ($Dy(NO_3)_3·6H_2O$), iron nitrate nonahydrate ($Fe(NO_3)_3·9H_2O$), chromium nitrate nonahydrate ($Cr(NO_3)_3 \cdot 9H_2O$), ethylene glycol ($C_2H_6O_2$), citric acid ($C_6H_8O_7$). First, all nitric acid compounds were mixed and dissolved in deionized water according to a certain ion ratio ($Dy^{3+}$: $Fe^{3+}$: $Cr^{3+}$ = 1: 0.1/0.9: 0.9/0.1). Then, the excess citric acid and a certain amount of ethylene glycol are mixed in a molar ratio of 1:1. Stir the solution uniformly on a magnetic stirrer at 80 °C until a gel form. Then, heat the gel to 120 °C. The resulting powder was pre-calcined at 600 °C for 12 h, then calcined at 1200 °C for 24 h, and cooled to obtain the final nanopowder sample.

X-ray diffraction measurements (XRD)

X-ray diffraction (XRD) experiments were carried out in Siemens D500 Cu Kα ($\lambda$ = 1.5418 Å) diffractometer with the range of 20° to 80° (rate of 0.02°/s). The obtained data were processed with Fullprof software.

Mössbauer spectrum Test

The transmission $^{57}$Fe Mössbauer spectrum of $DyFe_{1-x}Cr_xO_3$ ($x$=0.1 and 0.9) were collected at 300 K on SEE Co W304 Mössbauer spectrometer with a $^{57}$Co/Rh source in transmission geometry equipped in a cryostat (Advanced Research Systems, Inc.,4 K). The data results were fitted with MössWinn 4.0 software.

Magnetic Measurement

The thermomagnetic curves (M-T) of the $DyFe_{1-x}Cr_xO_3$ ($x$=0.1 and 0.9) were obtained in the external field of 100 Oe and the temperature in the range of 10 K to

400 K. And the magnetization curves (M-H) under -7 T to 7 T magnetic field were measured at 300 K temperature.

**Results and discussions**

XRD analysis

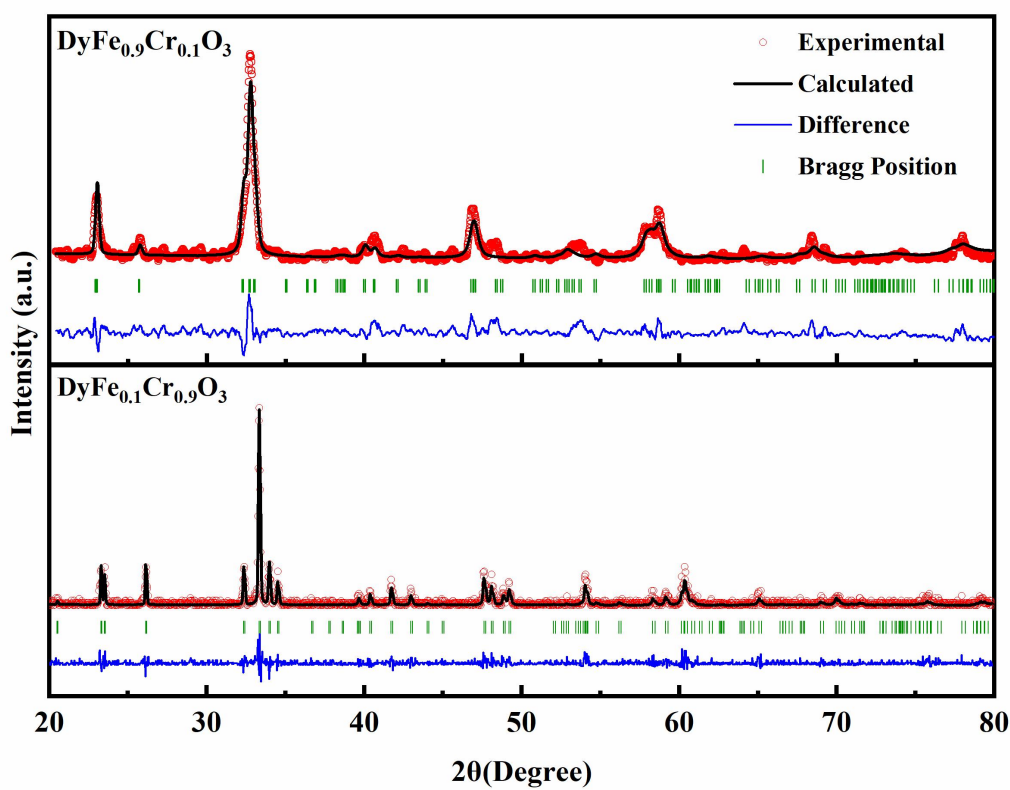

**Fig. 1.** The refined XRD patterns of DyFe$_{1-x}$Cr$_x$O$_3$ ($x$=0.1 and 0.9).

**Table. 1.** The refined lattice parameters and unit cell volume of DyFe$_{1-x}$Cr$_x$O$_3$ ($x$=0.1 and 0.9).

| Sample | $a$ (Å) | $b$ (Å) | $c$ (Å) | Cell Volume (Å$^3$) | ASD (%) |
|---|---|---|---|---|---|
| DyFe$_{0.9}$Cr$_{0.1}$O$_3$ | 5.5967 | 7.6334 | 5.3134 | 227.0006 | 0 |
| DyFe$_{0.1}$Cr$_{0.9}$O$_3$ | 5.2753 | 5.5310 | 7.5653 | 220.7387 | 0 |

Figure. 1 shows the XRD patterns of DyFe$_{1-x}$Cr$_x$O$_3$ ($x$=0.1 and 0.9) refined by Fullprof software. Both samples exhibited good crystallinity and only had a single-phase structure, in which DyFe$_{0.9}$Cr$_{0.1}$O$_3$ belonged to the orthorhombic *Pbnm* space group, and DyFe$_{0.1}$Cr$_{0.9}$O$_3$ belonged to the orthorhombic *Pnma* space group. It is not difficult to see that when the Cr doping amount increases, although the architecture maintains the orthogonal perovskite type, its orthogonal symmetry has changed greatly. The structural parameter ($D_{OD}$) is introduced to characterize the orthogonal distortion in the following formula: [20]

$$D_{OD} = \frac{1}{3}\sum_{i=1}^{3}\left|\frac{\alpha_i - \bar{\alpha}}{\bar{\alpha}}\right| * 100\% \#(1)$$

For the lattice of the *Pbnm* space group, $\alpha_1 = a$, $\alpha_2 = b$, $\alpha_3 = c/\sqrt{2}$, $\bar{\alpha} = (a*b*c/\sqrt{2})^{\frac{1}{3}}$. And for *Pnma* space group, $\alpha_1 = c$, $\alpha_2 = a$, $\alpha_3 = b/\sqrt{2}$, $\bar{\alpha} = (a*b*c/\sqrt{2})^{\frac{1}{3}}$.

The lattice constant, unit cell volume, and anti-site defect factor (ASD) are presented in Table 1. A rough ASD value can be obtained by Fullprof refinement, indicating the ratio of Fe/Cr ions on the B site to occupy each other. According to the refinement results of these two samples, we calculated that their ASD values are basically 0 %. This is because in the case of large ratio differences, the dopant ions will only occupy the lowest energy positions. A large ASD value occurs only when the B-site Fe/Cr ion ratios are close to or even the same, as in our previous study of NdFe$_{0.5}$Cr$_{0.5}$O$_3$. [9] As Cr ions far exceed Fe ions, both the lattice constant and the unit cell volume decrease accordingly. This is enough to see the effect of ionic radius on it.

[19, 21]

Mössbauer spectrum and Magnetization curve analysis

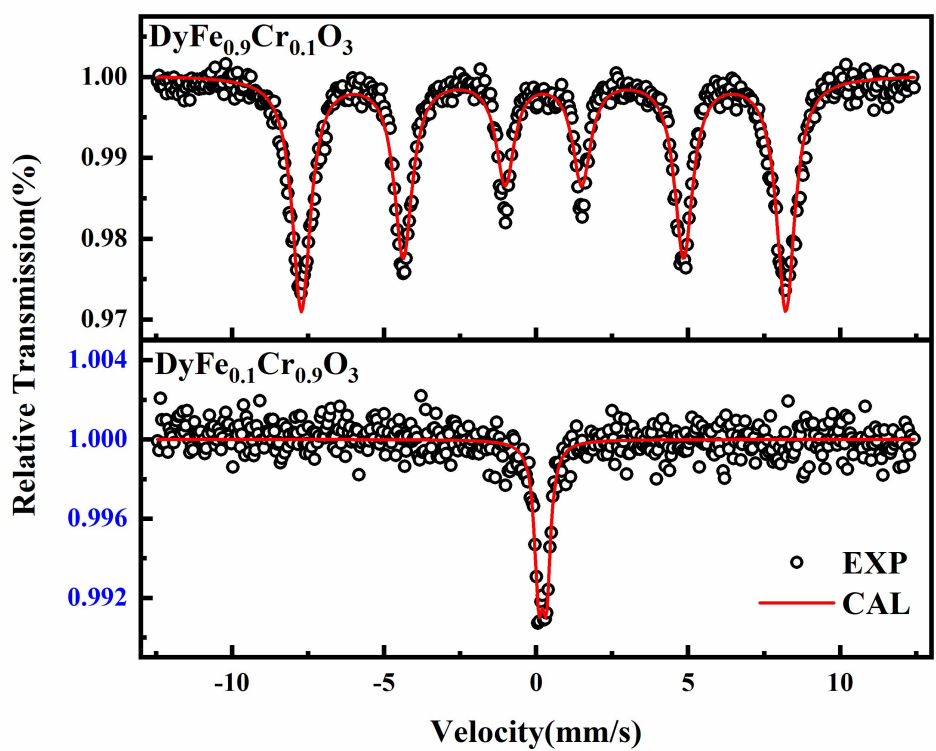

**Fig. 2.** Mössbauer spectrum of DyFe$_{1-x}$Cr$_x$O$_3$ (*x*=0.1 and 0.9) at 300 K.

**Table. 2.** Hyperfine parameters of DyFe$_{1-x}$Cr$_x$O$_3$ (*x*=0.1 and 0.9).

| Sample | IS (mm/s) | QS (mm/s) | H (T) | Γ (mm) | Fe Valence |
|---|---|---|---|---|---|
| DyFe$_{0.9}$Cr$_{0.1}$O$_3$ | 0.244 | 0.004 | 49.516 | 0.722 | Fe(III) HS |
| DyFe$_{0.1}$Cr$_{0.9}$O$_3$ | 0.216 | 0.256 | / | 0.330 | Fe(III) HS |

IS: isomer shift, QS: quadruple splitting/shift, Γ: Lorentzian linewidth, H: hyperfine magnetic field, HS: high spin.

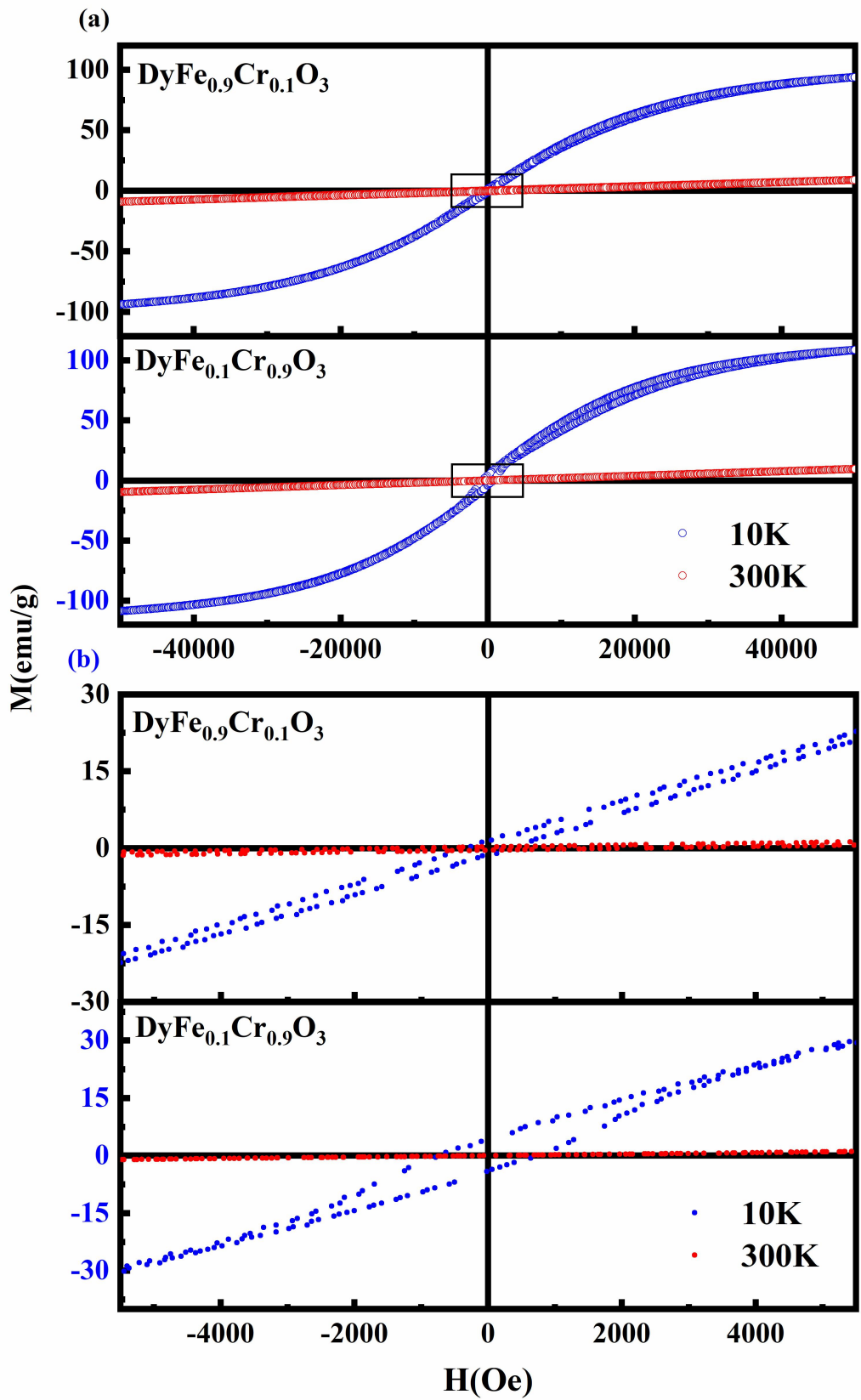

**Fig. 3.** **(a)** Magnetization curves of DyFe$_{1-x}$Cr$_x$O$_3$ ($x$=0.1 and 0.9) at 10 K and 300 K, respectively. **(b)** The enlarged view of the external magnetic field from -5500 Oe to 5500 Oe.

The Mössbauer spectra of DyFe$_{1-x}$Cr$_x$O$_3$ ($x$=0.1 and 0.9) are shown in Figure 2. The Mössbauer spectra at 300 K have obvious differences, which are similar to the general rare-earth iron-chromium perovskites. [22,23] In DFCO with high iron content, the sextet spectrum caused by the hyperfine field can prove its ferromagnetic or antiferromagnetic characteristics. In the samples with high chromium content, the expected doublet state is associated with macroscopic paramagnetism.

The hyperfine parameters fitted by MössWinn 4.0 software are shown in Table 2. In both samples we fit only one iron phase with good results. This also proves the homogeneity of the single phase in the DFCO system. By comparing isomer shift (IS) and quadruple splitting (QS), we believe that Fe ions are in the state of positive trivalent high spin, which will provide an important basis for our subsequent fitting work.

The measurement results of the magnetization curves of the samples are shown in Figure 3. On the whole, the two samples have very similar behaviors under the excitation of the external magnetic field, and the final magnetization is also relatively close. Distinct magnetization processes are observed at 10 K and 300 K, respectively. At 300 K, the magnetization curves of the two samples approached a straight line. The difference is that in the DyFe$_{0.9}$Cr$_{0.1}$O$_3$, a very fine hysteresis loop appears, indicating that an irreversible magnetic component appears in the DyFe$_{0.9}$Cr$_{0.1}$O$_3$. According to our previous research on $R$Fe$_{1-x}$Cr$_x$O$_3$ and reference to previous reports, we can understand it as the existence of a weak ferromagnetic phase. Between 300 K and Néel temperature, the spin angle that promotes the reduction of Fe$^{3+}$/Cr$^{3+}$ ions are slightly less than 180° due to the superexchange interaction caused by residual electrons and spin-orbit coupling, and a net magnetic moment appears. Weak ferromagnetic behavior is thus recorded below this temperature. [24] In DyFe$_{0.1}$Cr$_{0.9}$O$_3$, however, 300 K has not undergone the Néel phase transition, thus obtaining paramagnetic characteristics. At 10 K, both samples have undergone a Néel phase transition as well as an antiferromagnetic to weak ferromagnetic transition. Therefore, in the magnetization curve at 10 K, both have obvious hysteresis loops, which is the contribution of weak ferromagnetism. But as the temperature decreases, the paramagnetic contribution of Dy$^{3+}$-Dy$^{3+}$ (S=5/2) becomes very large, even enough to mask the weak ferromagnetic contribution of Fe$^{3+}$/Cr$^{3+}$. Especially when the external field is very large, it is difficult to observe the hysteresis loop for both samples, and both show the tendency of paramagnetic magnetization saturation.

Thermomagnetic analysis and Molecular Field Model Fitting

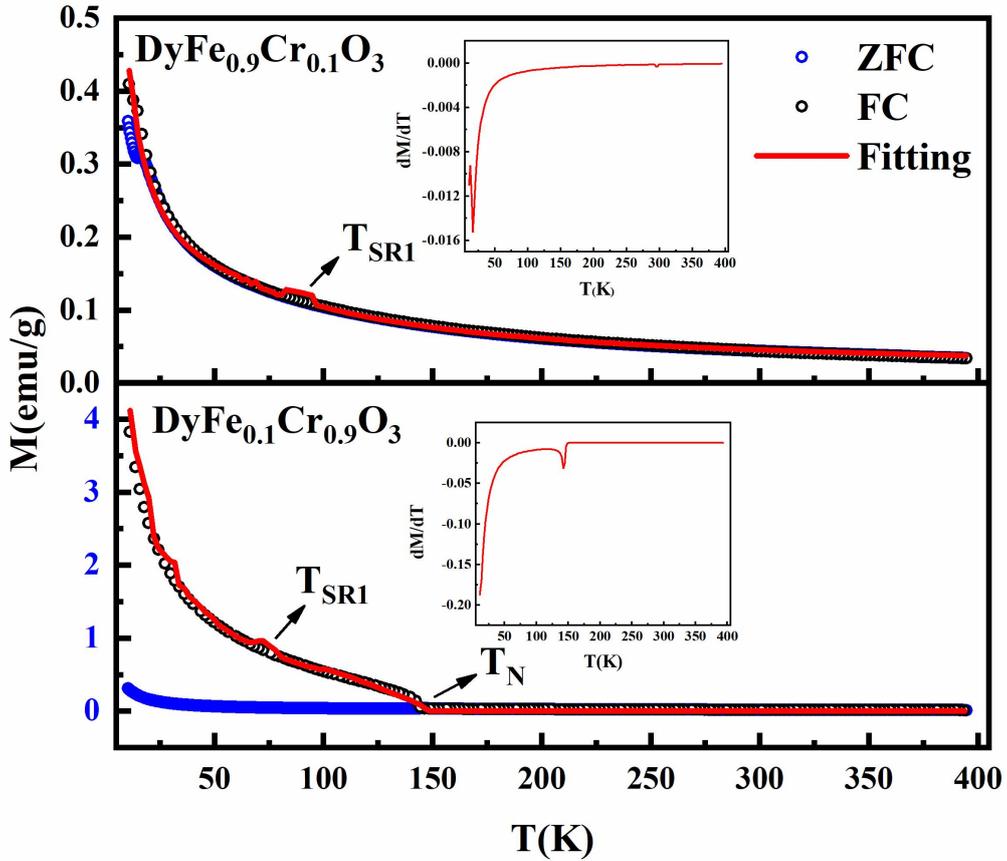

**Fig. 4.** Thermomagnetic curves and fitting results of DyFe$_{1-x}$Cr$_x$O$_3$ ($x$=0.1 and 0.9) in the temperature range from 10 K to 400 K. The inset shows the image of d$M$/d$T$ vs $T$.

The field cooling (FC) and zero field cooling (ZFC) curves of DFCO in the temperature range from 10 K to 400 K are shown in Figure. 4. In this temperature range, DFCO shows complex magnetic properties. On the whole, FC/ZFC showed an obvious upward trend with the decrease of temperature, but it is obvious that DyFe$_{0.1}$Cr$_{0.9}$O$_3$ can finally achieve a larger magnetization, that is to say, the ground state magnetization of DyFe$_{0.1}$Cr$_{0.9}$O$_3$ is much larger than that of DyFe$_{0.9}$Cr$_{0.1}$O$_3$. This originates from differences in exchange interactions in the DFCO system. Through the phase transition temperature of pure ferrite and pure chromite in molecular field theory, the exchange constants $J_{F-F}/J_{C-C}$ of $Fe^{3+}$-O-$Fe^{3+}$/$Cr^{3+}$-O-$Cr^{3+}$ can be directly calculated to be -25.8 K and -6 K, respectively. Therefore, the exchange that is strongly related to the low-temperature magnetization is not the exchange of $Fe^{3+}$-O-$Fe^{3+}$/$Cr^{3+}$-O-$Cr^{3+}$, but the exchange between B site ions and A being rare earth ions, that is, $Dy^{3+}$-O-$Fe^{3+}$/$Dy^{3+}$-O-$Cr^{3+}$.

To observe the magnetic phase transition of the two samples over the entire temperature range, we plotted the change in d$M$/d$T$ vs $T$ (Illustrations in Figure 4.). It is found that the $DyFe_{0.9}Cr_{0.1}O_3$ still maintains a weak magnetization in the whole temperature range, and no obvious Néel temperature point is observed. In $DyFe_{0.1}Cr_{0.9}O_3$, a very obvious magnetic transition occurs at about 147 K, and the magnetization is almost zero above this temperature, and the paramagnetic characteristics are very obvious. This also corresponds to the result of the Mössbauer spectrum. Two samples also have a very obvious magnetic phase transition point at very low temperature. Based on previous literature, this should not be the phase transition point where the $Dy^{3+}$ ion interacts with itself, but more likely the phase transition point where the interaction with the B-site ion occurs. This is because the antiferromagnetic phase transition of $Dy^{3+}$-$Dy^{3+}$ is at a much lower temperature (about 5 K).

In order to gain a more accurate understanding of the intrinsic origin of the magnetic properties of DFCO, a simple approximate simulation of the system is carried out using a four-sublattice molecular field model. In this process we consider only the exchange interactions in $Dy^{3+}$-$Fe^{3+}$, $Dy^{3+}$-$Cr^{3+}$, $Fe^{3+}$-$Fe^{3+}$, $Cr^{3+}$-$Cr^{3+}$, $Fe^{3+}$-$Cr^{3+}$, the reason for omitting the $Dy^{3+}$-$Dy^{3+}$ exchange interactions is that the paramagnetic contribution is very large over the whole simulated temperature range.

For $DyFe_{0.9}Cr_{0.1}O_3$, the mean field is expressed as:

$$H_{Dy} = \lambda_{DyCr}M_{Cr} + \lambda_{DyFe^b}M_{Fe^b} + \lambda_{DyFe^a}M_{Fe^a} + h \#(2)$$

$$H_{Fe^a} = \lambda_{Fe^aFe^b}M_{Fe^b} + \lambda_{Fe^aCr}M_{Cr} + \lambda_{Fe^aDy}M_{Dy} + h \#(3)$$

$$H_{Fe^b} = \lambda_{Fe^bFe^a}M_{Fe^a} + \lambda_{Fe^bDy}M_{Dy} + h \#(4)$$

$$H_{Cr} = \lambda_{CrFe^a}M_{Fe^a} + \lambda_{CrDy}M_{Dy} + h \#(5)$$

For $DyFe_{0.1}Cr_{0.9}O_3$, the mean field is expressed as:

$$H_{Dy} = \lambda_{DyFe}M_{Fe} + \lambda_{DyCr^b}M_{Cr^b} + \lambda_{DyCr^a}M_{Cr^a} + h \#(6)$$

$$H_{Cr^a} = \lambda_{Cr^aCr^b}M_{Cr^b} + \lambda_{Cr^aFe}M_{Fe} + \lambda_{Cr^aDy}M_{Dy} + h \#(7)$$

$$H_{Cr^b} = \lambda_{Cr^bCr^a}M_{Cr^a} + \lambda_{Cr^bDy}M_{Dy} + h \#(8)$$

$$H_{Fe} = \lambda_{FeCr^a}M_{Cr^a} + \lambda_{FeDy}M_{Dy} + h \#(9)$$

where $\lambda_{ij}$ represents the molecular field constant between sublattice $i$ and sublattice $j$, and it is positively related to the exchange constant $J_{ij}$, $h$ is the external magnetic field, and $M_i$ is the magnetisation intensity of sublattice $i$. $M_i$ can be expressed as:

$$M_i = \chi_i N_A g \mu_B S_i B_{S_i}\left(\frac{g\mu_B S_i H_i}{k_B T}\right) \#(10)$$

Where $\chi_i$ is the molar quantity of $i$ ions, $g$ is the lande factor, and $\mu_B$ represents the Bohr magneton. $N_A$ is the Avogadro constant. $S_i$ is the spin quantum number of $i$ ions ($S_{Fe}$ = 5/2, $S_{Cr}$ = 3/2, $S_{Dy}$ = 5/2). The exchange constant $J_{MM}$ between $M$ and $M$

ions can be calculated by:

$$|J_{MM}| = \frac{2Z_{MM}S_M(S_M + 1)}{3k_B T_N^M} \#(11)$$

With this equation we can understand that once the corresponding Néel temperatures for rare-earth ferrites and chromium oxides are obtained, we can calculate the corresponding approximate exchange constants. Based on previous studies, we find that the phase transition temperatures of $DyFeO_3$ and $DyCrO_3$ are approximately 645 K and 150 K, respectively, with corresponding exchange constants of -25.8 K and -6 K for $J_{F-F}$ and $J_{C-C}$, respectively. Thus the magnetisation intensity at each temperature point can be obtained by means of equations (2) - (10). The algorithm used for the fitting is the latest Marine Predator Algorithm (MPA). [24]

At the same time, we define the effective spin of the A-site rare earth ion as the projection of its total spin onto the B-site spin plane. This is different from the research methods of many previous scholars. Before this, most researchers generally defined the spin of A-site rare earth ions as a constant, and did not think it was changing. However, during the fitting process, we found that in addition to the spin reorientation of the $Fe^{3+}/Cr^{3+}$ ions at the B site, the spins of the A-site ions also changed at the same time, and the final error was very small. This work has been well validated in Nd, Pr, Er rare earth perovskites. In the high temperature section, the influence of the spin of rare earth ions on the magnetization is almost negligible, so the high temperature section is selected to first fit to obtain the relevant exchange constants (in Table. 3). The obtained exchange constants are then used to fit the A-site effective spins over the entire temperature range. The thermomagnetic curve after junction fitting is shown as the red line in Figure. 4.

It can be seen that the fitting is very good, and the error can even be less than $10^{-3}$ at some temperature points, and only slightly larger errors appear in some temperature ranges. Around 75 K, the fitted curves for both samples exhibited slightly larger errors. In many literatures, [7,12, 22, 25,26] 75 K is a very specific temperature point, the time at which the B-site spin reorientation occurs. In the original data, the spin of the $Dy^{3+}$ ion will have a huge effect on the whole system due to the decrease in temperature, i.e. a very large paramagnetic shielding. In the experimental data, the paramagnetic effect far exceeds the contribution of weak ferromagnetism, so it is difficult to observe the magnetization change due to the B-site spin reorientation. In the fitting process, this detail is clearly manifested. In addition, we found that in the ZFC curve of $DyFe_{0.9}Cr_{0.1}O_3$, there is an obvious turning point at about 15 K, which is probably the phase transition temperature point where $Dy^{3+}$ ions are coupled with $Fe^{3+}/Cr^{3+}$ ions. While in $DyFe_{0.1}Cr_{0.9}O_3$, this process may be below 10 K without being observed.

The fitted effective spin of $Dy^{3+}$ ions as a function of temperature is shown in Figure. 5. The corresponding $Dy^{3+}$ spin direction and the angle between the $Fe^{3+}/Cr^{3+}$ spin plane is also reflected.

**Table. 3.** Exchange constant of $DyFe_{1-x}Cr_xO_3$ ($x$=0.1 and 0.9).

| Sample | $J_{F-F}$ | $J_{C-C}$ | $J_{F-C}$ | $J_{D-F}$ | $J_{D-C}$ |
| --- | --- | --- | --- | --- | --- |
| $DyFe_{0.9}Cr_{0.1}O_3$ | -25.800 | -6.000 | -12.214 | -5.119 | -11.385 |
| DyFe$_{0.1}$Cr$_{0.9}$O$_3$ | -25.800 | -6.000 | -10.439 | -2.005 | -5.537 |

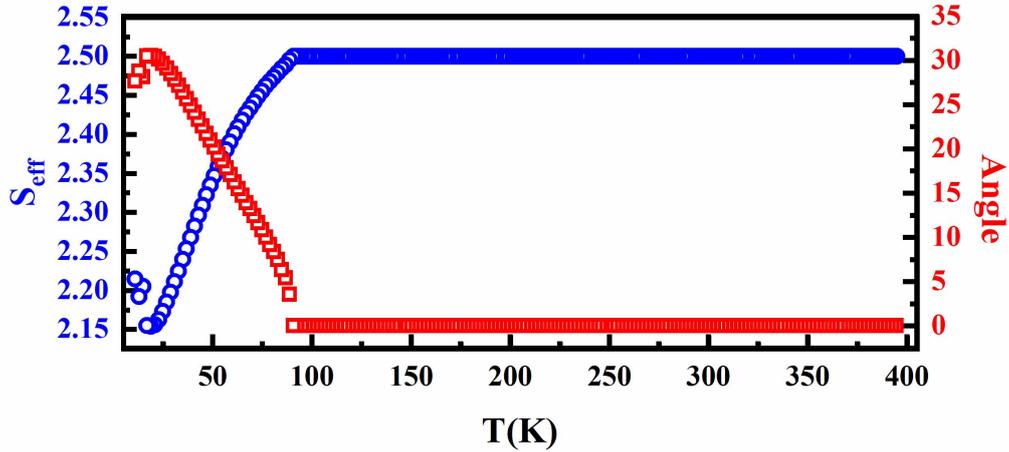

**Fig. 5.** The effective spins of Dy$^{3+}$ ions and the corresponding angle with the Fe$^{3+}$/Cr$^{3+}$ spin plane change with temperature.

**Conclusions**

In summary, Mössbauer spectroscopy and magnetic measurements were performed on single-phase DyFe$_{0.9}$Cr$_{0.1}$O$_3$ and DyFe$_{0.1}$Cr$_{0.9}$O$_3$ prepared by a simple sol-gel method. The Mössbauer spectra at room temperature are consistent with the results shown by the magnetization curves, showing weak ferromagnetic properties in DyFe$_{0.9}$Cr$_{0.1}$O$_3$ and obvious paramagnetic behavior in DyFe$_{0.1}$Cr$_{0.9}$O$_3$. At 10 K, both samples have crossed the antiferromagnetic to weak ferromagnetic transition temperature, but the magnetization curves will tend to paramagnetic saturation due to the action of high-spin Dy$^{3+}$ ions under a large external magnetic field.

By fitting the molecular field model of the four-sublattice and MPA algorithm, the thermomagnetic curve with a high degree of fit is obtained, and the relevant exchange constants are obtained. Different from past theories, we define the effective spin of rare earth ions at the A site as the projection of its full spin on the spin plane at the B site, and fit it as an unknown parameter over the entire temperature range, and finally get the effective spin about temperature change relationship.

**Conflicts of interest**

There are no conflicts to declare.

**Data Availability**

The datasets generated and/or analysed during the current study are available from the corresponding author on reasonable request.

**Acknowledgements**

This work was supported by National Natural Science Foundation of China (grant number 12105137, 62004143), the Central Government Guided Local Science and Technology Development Special Fund Project (2020ZYYD033), the National Undergraduate Innovation and Entrepreneurship Training Program Support Projects of China, the Natural Science Foundation of Hunan Province, China (grant number S202110555177), the Natural Science Foundation of Hunan Province, China (grant number 2020JJ4517), Research Foundation of Education Bureau of Hunan Province, China (grant number 19A433, 19C1621).